\documentclass[fleqn]{article}
\usepackage{gc,amsfonts}

\def\mn{_{\mu\nu}}
\def\MN{^{\mu\nu}}
\def\mN{_\mu^\nu}

\def\uu#1{{\mathop{u}\limits_#1}}
\def\s{_{s\sigma}}
\def\S{^{s\sigma}}

\def\Mab{\lims{M}_{(ab)}}
\def\mab{\lims{\mu}_{(ab)}}
\def\MO{\limr{1.5pt}{M}{}{(0)}}
\def\mO{\lims{\mu}_{(0)}}
\def\Pab{\lims{\cal P}_{(ab)}{}}
\def\zab{\lims{\zeta}_{(ab)}{}\nhq}
\def\Zab{\lims{Z}_{(ab)}\nhq}
\def\ZO{\lims{Z}_{(0)}{\nhq\,}}
\def\zO{\lims{\zeta}_{(0)}{\nhq\,}}
\def\sO{s_{{}_0}}

\eqsection

\heads  {K.A. Bronnikov and E.A. Tagirov}
        {Quantum theory of scalar field in isotropic world}

\begin{document}
\twocolumn[
\prepno{gr-qc/0412138}{\GC{10} 249 (2004)}

\vspace*{1cm}
\rightline{\Large \sf RETRO}
\vspace*{.4cm}

\rightline{\parbox{14cm}{\sf In this section of the journal we continue
publishing in English some important papers published earlier in Russian and
unknown to the English-speaking scientific community.}}

\vspace*{2cm}

\Title {Quantum theory of scalar field in isotropic world\foom 1}

\Authors{K.A. Bronnikov\foom 2} {and E.A. Tagirov\foom 3}
   {VNIIMS, 3-1 M. Ulyanovoy St., Moscow 119313, Russia;\\
    Institute of Gravitation and Cosmology, PFUR,
        6 Miklukho-Maklaya St., Moscow 117198, Russia}
   {Joint Institute for Nuclear Research, Dubna 141980, Russia}

\Abstract
   {Fock representations are constructed for a free scalar field
   in the closed and quasi-Euclidean isotropic cosmological models.
   Invariance of their cyclic vector (vacuum) under isometries and the
   correspondence principle single out a class of unitarily equivalent
   representations.}

] 
\foox 1 {Published for the first time as
     {\it Preprint JINR\/} P2-4151, Dubna, 1968.}
\email 2 {kb@rgs.mccme.ru; \ kb20@yandex.ru}
\email 3 {tagirov@thsun1.jinr.ru}

\section{Introduction}

   There is a certain hope that an analysis of applicability of the basic
   notions and methods of quantum field theory (QFT) in a space-time
   manifold which differs from the Minkowski world can shed light upon some
   problems of this theory as a whole. On the other hand, the idea has been
   repeatedly expressed [1--3] that the large-scale geometry of space-time
   can determine certain (or to a certain extent) the properties of the
   micro-world. The geometry of the Universe is apparently essentially
   non-pseudo-Euclidean, and, as long as QFT is adequate to the dynamics of
   elementary processes, it is natural to seek manifestations of a
   fundamental relationship (if any) between cosmology and the micro-world
   in the QFT structure.

   It is from these positions that, as the first step, a study of quantum
   theory of a free scalar field in the de Sitter world was undertaken [4,
   5]. It was shown there, in particular, that the notion of a free
   particle, which is commonly related to irreducible representations of the
   Poincar\'e group, can be related to the de Sitter group in a similar
   manner only if one invokes some particular formulation of the
   correspondence principle. In the present paper we make an attempt to
   extend the methods and results of Ref.\,[5] to the general case of
   isotropic models of the Universe. We consider in detail closed models
   [the group of isometries $O(4)$] and at the end give the modifications of
   the main results for quasi-Euclidean models [the group of isometries
   $O(3) \times T_3$].

   A closed isotropic Universe $F_4$ may be geometrically reproduced as the
   rotation hypersurface
\bearr
   x^a x^a = (x^1)^2 + (x^2)^2 + (x^3)^2 + (x^4)^2
            	= \rho^2 f^2 \biggl(\frac{x^0}{\rho}\biggr),
\nnn \cm
    f > 0, \cm |f'| < 1                                \label{1.1}
\ear
   in the five-dimensional pseudo-Euclidean space-time $E_5$ with the metric
\[
     ds^2 = (dx^0)^2 - dx^a \, dx^a.
\]
   It is convenient to choose the parameter $\rho$ in such a way that
   $f(0) = 1$.

   In the coordinates $y^\mu$ ($y^0 \equiv \eta$) determined by the relations
\bearr
    x^0 = \rho \int_{0}^{\eta} \sqrt{b^2(\eta') +
                \dot{b}(\eta')}\, d\eta',
\nnn
    x^a = \rho\, b(\eta)\, n^a(y^1,\ y^2,\ y^3); \cm n^a n^a =1,
\earn
   where $b(\eta) = f(x^0/\rho)$ and $\dot{b}(\eta) = db/d\eta$, the metric
   of $F_4$ has the form
\beq
     ds^2 = g\mn dy^\mu dy^\nu = \rho^2 b^2(\eta)
            (d\eta^2 - h_{ij} dy^i dy^j)             \label{1.2}
\eeq
   where $h_{ij} = \d_i n^a \d_j n ^a$, $\d_i = \d/\d y^i$, Greek indices
   take the values from 0 to 3, Latin ones $i,j,k$ from 1 to 3 and $a,b$
   from 1 to 4. If the metric is specified in the form (1.2), then one can
   obtain \eq (1.1) by putting
\[
    \eta = \int_{0}^{x^0/\rho} \sqrt{1 - [f'(\alpha)]^2}
            \frac{d \alpha}{f(\alpha)}.
\]

   De Sitter space corresponds to $b(\eta) = 1/\cos \eta$; for the Friedmann
   models with $p = \eps/3$ and $p=0$ ($\eps$ = energy density and $p$ =
   pressure) one has $b(\eta) = \cos \eta$ and $b(\eta) = \cos^2 (\eta/2)$,
   respectively [6].

\section{The field equation}

    As follows from [7] and [5], the scalar field equation in a Riemannian
    space has the form
\beq
       (\DAL + m^2 + R/6) \phi = 0 \cm (c = \hbar =1),       \label{2.1}
\eeq
    where $R = g\MN R\mn$ is the scalar curvature,
    $R\mn = R^\alpha{}_{\mu \alpha \nu}$ is the Ricci tensor, and the sign
    of the curvature tensor is chosen so that
\[
    \nabla_\mu \nabla_\nu A_\sigma - \nabla_\nu \nabla_\mu A_\sigma
        = R^\alpha{}_{\sigma\mu\nu} A_\alpha
\]
    for any vector $A_\alpha$; $\DAL = \nabla^\alpha \nabla_\alpha$ is the
    d'Alembert operator, and $\nabla_\alpha$ is a covariant derivative.

    \eq (2.1) is obtained by variation with respect to $\phi$ of
    the following action integral:
\bearr
    A = \int L \sqrt{-g} d^4 y                            \label{2.2}
\nnn
    = \Half \int \biggl[g\MN \d_\mu\phi\,\d_\nu\phi -
       		\biggl(m^2 + \frac{R}{6}\biggr)\phi^2\biggr]\sqrt{-g} d^4 y,
\ear
    where $d^4 y = dy^0 dy^1 dy^2 dy^3$. Varying the integral (2.2) with
    respect to $g\MN$, we obtain the (metric) energy-momentum tensor (EMT)
\beq
     T\mn = T^{\rm can}\mn - \frac{1}{6}                      \label{2.3}
            (R\mn + \nabla_\mu \nabla_\nu - g\mn \DAL) \phi^2,
\eeq
    where $T^{\rm can}\mn$ is the canonical EMT,
\[
    T^{\rm can}\mn = \d_\mu \phi \d_\nu \phi - g\mn L.
\]
    The tensor (2.3) possesses the properties
\[
    T\mn = T_{\nu\mu}, \cm T^\alpha_\alpha =m^2 \phi^2,
            \cm \nabla^\alpha T_{\alpha\mu} =0.
\]

    Let us pass to quantum field theory by specifying commutation relations
    on a certain spacelike hypersurface $\Sigma$ in $F_4$:
\bearr
    [\phi(M),\ \phi(M')] =0,                        \label{2.4}
\cm
    [\d_\mu\phi(M),\ \d_\nu\phi(M')] =0,
\nnn
    \intl_\Sigma [\phi(M),\ \d_\mu\phi(M')] f(M') d\sigma^\mu(M')
            = i f(M),
\ear
    where $M,\ M' \in \Sigma$, $f(M)$ is an arbitrary function and
    $d\sigma^\mu (M)$ is the vector area element on $\Sigma$. In the
    Heisenberg picture (which is fixed by the choice of $\Sigma$), the
    operator $\phi$ satisfies \eq (2.1) and the initial data (2.4).
    Let us choose as $\Sigma$ the surface $\eta=0$, then
    $d\sigma^\mu = \delta^{\mu 0} \rho^2 \sqrt{h} d^3 y$.

    In $F_4$, due to (1.2), \eq (2.1) has the form
\[
     \biggl[ \frac{1}{b^2} \frac{\d}{\d\eta} b^2 \frac{\d}{\d\eta}
    -\Delta  + m^2 \rho^2 b^2 + \frac{b + \ddot b}{b}\biggr]\phi =0,
\]
    where $\Delta = h^{-1/2} \d_i \bigl( h^{1/2} h^{ik} \d_k\bigr)$ is the
    Laplace operator on the sphere $n^a n^a =1$. Separating the variables,
    we obtain $\phi$ as an expansion in harmonic polynomials on the sphere
    $n^a n^a =1$:
\bearr
    \phi(\eta,y) = \frac{1}{b(\eta)\rho} \sum_{s=0}^{\infty}
           \sum_{\sigma=1}^{(s+1)^2}
           u_{s\sigma} (\eta) P^{s\sigma} (y),        \label{2.5}
\yyy
    [\Delta + s(s+2)] P^{s\sigma} (y) = 0,                    \label{2.6}
\nnnv
    P^{s\sigma} (y) = \frac{\sqrt{2^s (s+1)}}{\pi\sqrt{2}}
            P^\sigma_{a_1\ldots a_s}n^{a_1}\ldots n^{a_s},
\ear
    where the tensors $P^\sigma_{a_1\ldots a_s}n^{a_1}\ldots n^{a_s}$, which
    are symmetric over all $a_i$ and have a zero trace over a pair of
    indices, are orthonormalized:
\[
    P^\sigma_{a_1\ldots a_s} P^{\sigma'}_{a_1\ldots a_s}
    = \delta^{\sigma\sigma'}, \cm
    \sigma,\ \sigma' = 1, \ldots, (s+1)^2.
\]
    Here and henceforth $y = (y^1,\ y^2,\ y^3)$.

    The operator $u_{s\sigma}(\eta)$ evidently obeys the equation
\beq     \wide
    \ddot u + [(s+1)^2 + m^2 \rho^2 b^2(\eta)] u =0.       \label{2.7}
\eeq
    Let us introduce its two linearly independent solutions $u^{\pm}$,
    specifying them by the initial conditions
\bearr
          u^{\pm}_s(0) = \frac{1}{\sqrt{\sO}},
\cm
         \dot u^{\pm}_s(0) = \pm \sqrt{\sO},
\nnn
     \sO = \sqrt{(s+1)^2 + m^2 \rho^2}.                    \label{2.8}
\ear

    The Cauchy problem (2.7), (2.8) is equivalent to the Volterra integral
    equation
\bearr
    u_s^{\pm} (\eta) = w^{\pm} (\sO,\eta)                  \label{2.9}
\nnn\qquad
    + 2 m^2\rho^2 \intl_{0}^{\eta}[1-b^2(\eta')]\, g(\eta,\eta')
        u_s^{\pm}(\eta') d\eta',
\ear
    where the functions
\beq                                                           \label{2.10}
     w^{\pm} (\sO,\eta) = \frac{1}{\sqrt{\sO}} \e^{\pm i \sO\eta}
\eeq
    are solutions of the equation
\beq
    \ddot w + \sO^2\, w =0                                 \label{2.11}
\eeq
    under the initial conditions (2.8); the function
\bearr
    g(\eta,\eta') = - \sign (\eta-\eta')
\nnn \ \
    \times
    \left| \matrix{ w_1(\eta')  &  w_2(\eta') \cr
                  \dot w_1(\eta')  &  \dot w_2(\eta') \cr} \right|^{-1}\
    \left| \matrix{ w_1(\eta')  &  w_2(\eta') \cr
                   w_1(\eta)   &  w_2(\eta)  \cr}\right|
\nnn \inch
    = \frac{\sign(\eta-\eta')}{2\sO} \sin [\sO(\eta-\eta')]
\earn
    is a fundamental solution of \eq (2.11); $w_1$ and $w_2$ are its any two
    linearly independent solutions. Evidently,
\bear
    \Bigl(u_s^+(\eta)\Bigr)^* \eql u_s^-(\eta),
\nn
     W(u_s^+,\ u_s^-) \eql u_s^+\, \dot u_s^- - \dot u_s^+\, u_s^- = -2i.
\earn
    It follows from \eq (2.9) that the function $u_s^{\pm} (\eta)$ may be
    approximated with any precision by members of the uniformly convergent
    sequence of iterations:
\bearr
     \uu{0}{}_s^{\pm} (\eta) = w^{\pm} (\sO, \eta),            \label{2.12}
\nnn
    .\ .\ .\ .\ .\ .\ .\ .\ .
\nnn
     \uu{{n+1}}{}_s^{\pm}(\eta) = w^{\pm} (\sO, \eta)
\nnn \qquad
      +2m^2\rho^2 \int_0^{\eta} [1-b^2(\eta)] g(\eta,\eta')
            \uu{n}{}_s^{\pm} (\eta')\, d\eta'.
\ear
    This method is convenient for finding asymptotic expressions for
    $s\gg m\rho$. It is, in particular, easy to verify that
\beq
     u^+_s\, u^-_s = \frac{1}{s+1}                           \label{2.13}
    \biggl[ 1 - \frac{m^2\rho^2 b^2}{2(s+1)^2}
        + o\biggl(\frac{m^2\rho^2}{s^2} \biggr) \biggr].
\eeq

    It is clear that for $m=0$
\[
      u_s^{\pm} (\eta) = w^{\pm}(s+1,\eta) =
    \frac{1}{\sqrt{s+1}} \e^{\pm i(s+1)\eta}.
\]

    The operator $u_{s\sigma}(\eta)$ in (2.5) may be presented in the form
\bearr
      u_{s\sigma} (\eta) = \frac{\sqrt{\sO}}{2}             \label{2.14}
            q_{s\sigma} (u_s^+ + u_s^-)
       + \frac{i}{2\sqrt{\sO}} p_{s\sigma} (u_s^- - u_s^+),
\nnn
\ear
    where $q\s = u\s(0)$ and $p\s = \dot u\s(0)$. Owing to (2.5)
    and to the completeness of the
    system of harmonic polynomials on the sphere $n^a n^a =1$, one has
\bear
     u\s(\eta) \eql \rho b(\eta) \int \phi(\eta,y) P\S(y) d\sigma(y),
\nn
    \dot u\s(\eta) \eql \rho b(\eta) \int
    \biggl[\d_0 \phi(\eta,y) + \frac{\dot b}{b} \phi(\eta,y)\biggr]
            P\S(y) \, d\sigma.
\earn
    It follows from (2.4) that
\bearr
      [q\s,\ q_{s'\sigma'}] =0, \cm                        \label{2.15}
      [p\s,\ p_{s'\sigma'}] =0,
\nnn
      [q\s,\ p_{s'\sigma'}] = i \delta_{ss'} \delta_{\sigma\sigma'}.
\ear
    Using these relations, the expressions (2.5) and (2.14) and the harmonic
    polynomials addition theorem [8], one can present the commutator of
    field operators $\phi(M)$ taken at two arbitrary points $M_1$ and $M_2$
    in $F_4$ as the following series:
\bearr
     D(\eta_1,y_1;\ \eta_2,y_2) = i [\phi(M_1),\ \phi(M_2)]
\nnn \nhq\!
     = \frac{i}{2b(\eta_1)b(\eta_2)} \sum_{s=0}^{\infty}
       \frac{s+1}{2\pi^2} C_s^1 \Bigl(n^a(y_1)n^a(y_2)\Bigr)
        D_s (\eta_1,\, \eta_2),
\earn
    where $C_s^1$ is a Gegenbauer polynomial and
\[
     D_s (\eta_1,\ \eta_2) = \frac{1}{i}
        \Bigl[u_s^-(\eta_1)u_s^+(\eta_2)
             -u_s^+(\eta_1)u_s^-(\eta_2)\Bigr].
\]
    It is well known that knowledge of the commutation function $D$ makes it
    possible to solve the Cauchy problem for \eq (2.1) with initial data on
    an arbitrary surface $\Sigma$.

    It is of interest to note that, according to [5], the function $D$ in
    de Sitter space in case $m=0$ is concentrated on the light cone. On the
    other hand, any $F_4$ is conformal to de Sitter space, and, since
    conformal mappings preserve the light cone, one can evidently assert
    that $D$ for $m=0$ is also concentrated on the light cone in any $F_4$.
    This property, which is natural for a field with zero rest mass, will
    only take place if the field equation is chosen in the form (2.1).

\section{Fock representations}

    The next step in quantization is to construct a representation  of
    the commutation relations. To this end, we introduce the operators [9]
\beq \nq\,
     z^-\s = \frac{i}{\sqrt{2}}                              \label{3.1}
        \biggl(p\s \!-\sum_{t, \tau} T\s{}_{,t\tau} q_{t\tau}\biggr),
    \quad\   z^+\s = (s^-\s)^*,
\eeq
    where $T = S + iQ$ is an arbitrary symmetric matrix
    ($T\s{}_{,t\tau} = T_{t\tau,}{}\s$) with a positive-definite imaginary
    part $Q$. From (2.15) it follows
\[
      [z^+\s,\ z^-\s] = 0, \cm [s^-\s,\ z^+_{t\tau}] = Q\s{}_{,t\tau}.
\]
    Reversing the equality (3.1), we have
\def\zz{{\widetilde{\widetilde z}}}
\def\vac {\mbox{$|0\rangle\ $}}
\bear
    q\s \eql \frac{1}{\sqrt{2}}\biggl(\zz{}^-\s + \zz{}^+\s\biggr),
\nn
    p\s \eql \frac{1}{\sqrt{2}}\biggl(                     \label{3.2}
               T^*_{s\sigma,t\tau}\zz{}z^-_{t\tau}
            + T_{s\sigma,t\tau}\zz{}^+_{t\tau}\biggr),
\ear
    where
\[
    \zz{}^{\pm}\s = \sum_{t\tau} Q_{s\sigma,t\tau}^{-1} z^{\pm}_{t\tau}.
\]
    The state vector \vac, defined by the relations
\[
     z^-\s \vac =0, \cm \langle 0 \vac = 1,
\]
    will be called a quasi-vacuum. For each $T$, the states
\[
    |s_1 \sigma_1\ldots s_N \sigma_N\rangle
          = z^+_{s_1\sigma_1} \ldots z^+_{s_N \sigma_N} \vac
\]
    ($N$-quasiparticle states) form a basis of a certain Fock
    representation. Further, we will seek among these
    representations the one in which $|s_1\sigma_1\ldots s_N\sigma_N\rangle$
    might be regarded as a state with a certain number $N$ of particles.

\section{Conserved quantities and invariant quasi-vacuum states}

    It is, above all, evident that the quasi-vacuum of such a representation
    should be invariant under the group of isometries of the space-time $F_4$.
    It is, as is easily seen, the group $O(4)$ with the six generators
\[
     \Zab = i\zab^\alpha \d_{\alpha},\qquad
     \zab^\alpha = \delta^\alpha_j
            \Bigl(n^b\d_i n^a - n^a\d_i n^b\Bigr)h^{ij},
\]
    corresponding to rotations in the $(ab)$ planes of the embedding space
    $E_5$. Each Killing vector $\zab^\alpha$ determines a conserved (i.e.,
    independent of the choice of the spacelike hypersurface $\Sigma$)
    quantity
\[
     \Mab = \intl_\Sigma \zab^\alpha\, T_{\alpha\beta} d\sigma^\beta.
\]
    Calculating these integrals according to [5], we obtain
\[
     \Mab = \sum_{s,\sigma,\tau} (s+1) q_{s+1,\sigma}\
            \Pab_{s+1}^{\sigma\tau}\, p_{s+1,\tau},
\]
    where $\Pab_{s+1}$ are antisymmetric matrices with the elements
\[
    \Pab^{\sigma\tau}_{s+1} =
     P^{\sigma}_{a_1...a_s a} P^{\tau}_{a_1...a_s b} -
                P^{\tau}_{a_1...a_s a} P^{\sigma}_{a_1...a_s b}.
\]
    The requirement of invariance of a quasi-vacuum under isometries,
    expressed by the condition
\[
      \Mab \vac = \mab \vac,
\]
    where the $\mu$'s are c-numbers, leads to the equalities
\[
     T_{s\sigma,t\tau} = \delta_{st} \delta_{\tau\sigma} T_s,
    \cm   \mab =0,
\]
    with $T_s = S_s + iQ_s$, $Q_s >0$ being arbitrary numbers.

    Let us introduce new parameters $\lambda_s$ instead of $T_s$:
\[
     T_s = i\sO \frac{1-\lambda_s}{1+\lambda_s}, \cm |\lambda_s| < 1,
\]
    and form the new operators
\[
     c\s^- = \frac{1+\lambda_s}{\sqrt{1-|\lambda_s|^2}}
                \frac{z\s^-}{\sqrt{\sO}},
     \cm     c\s^+ = \left( c\s^-\right)^*,
\]
    obeying the canonical computation relations
\[
    [c\s^+,\ c_{t\tau}^+] = 0, \cm
    [c\s^-,\ c_{t\tau}^+] = \delta_{st} \delta_{\sigma\tau}.
\]
    It is natural to call them quasi-particle creation and annihilation
    operators. The expressions (3.2) are considerably simplified:
\bear \nq
     q\s \eql \frac{1}{\sqrt{2\sO (1+|\lambda_s|^2}}
           \left[(1-\lambda_s^*)c\s^- + (1+\lambda_s)c\s^+\right],
\nnv  \nq
     p\s \eql -i\sqrt{\frac{\sO}{2}}\frac{1}{\sqrt{1-|\lambda_s|^2}}
           \left[(1-\lambda_s^*)c\s^- + (1-\lambda_s)c\s^+\right],
\earn
    and we finally obtain for $\Mab$:
\[
     \Mab = i\sum_{s,\sigma,\tau} s c\s^+ \Pab_s^{\sigma\tau} c^-_{s\tau}.
\]

    Thus the requirement that the quasi-vacuum should be invariant under
    isometries selects a class of Fock representations, each of them being
    specified by a certain choice of the sequence of parameters
    $\{\lambda_s\}$.

    Furthermore, $F_4$ admits conformal transformations, among which it is
    sufficient for us to consider the one-parameter subgroup determined by
    the generator  $\ZO = \zO^{\alpha} \d_\alpha$, where the vector
    $\zO^\alpha = \delta^{\alpha 0}$ satisfies the generalized Killing
    equation
\[
       \nabla_\mu \zeta_\nu + n_\nu \zeta_\mu = 2\,\frac{\dot b}{b}\, g\mn.
\]
    Since $T^\alpha_\alpha =0$ for the massless scalar field, $m=0$,
    the integral
\[
    \MO = \intl_\Sigma \zO^\alpha T_{\alpha\beta} d\sigma^\beta
\]
    is conserved. The quasi-vacuum is invariant under these transformations,
\beq                            \label{4.2}
    \MO \vac = \mO \vac,
\eeq
    only if $\lambda_s =0$. Thus for $m=0$ we obtain a single Fock
    representation with an invariant quasi-vacuum, which therefore may be
    called the vacuum. In this case,
\[
     \MO = \Half \sum_{s,\sigma} (s+1)
            \left( c\s^- c\s^+ + c\s^+ c\s^- \right).
\]

    Lastly, in the general case, the relations (2.5), (2.14) and (4.1) lead
    to the following expression for the field operator:
\bearr
     \phi(\eta, y) = \frac{1}{\sqrt{2}b(\eta)\rho}             \label{4.3}
    \sum_{s=0}^{\infty} \sum_{\sigma=1}^{(s+1)^2} P\S (y)
\nnn  \cm\cm
        \times \left\{u_s(\eta) c\s^+  + u^*_s (\eta) c\s^-\right\},
\ear
    where
\[
     u_s(\eta) = \frac{u_s^+ + \lambda_s u_s^-}{\sqrt{1 - |\lambda_s|^2}}.
\]

\section{The correspondence principle: the quasi-classical limit}

    Now we will consider the restrictions on the sequence $\{\lambda_s\}$
    following from the correspondence principle in the following
    formulation. If the amplitudes
\beq                                                         \label{5.1}
    \psi\s (\eta,y) = \langle|\phi(\eta,y) c\s^+\vac
            = \frac{u_s^*(\eta) P\S(y)}{\sqrt{2}\rho\,b(y)}
\eeq
    are relativistic wave functions of a particle, they must be
    quasi-classical at large $s$. Of course, this is only a necessary
    condition following from the interpretation of the quantum number $s^2$
    as the spatial momentum squared [5]. It means that at large $s$ there
    must be such functions among $\psi_s = \sqrt{\alpha} \e^{iS}$ that
    $\alpha = |\psi_s |^2$ and $S = \arg \psi_s$ satisfy the
    Hamilton-Jacobi equation
\[
    g\MN \d_\mu S\, \d_\nu S = m^2
\]
\onecol \noi
    and the continuity equation
\[
    g\MN \nabla_\mu (\alpha\, \d_\nu S) =0
\]
    (see more details in [5]). Separating the variables,
\[
    \alpha = A(\eta) C(y), \cm S = T(\eta) +  U(y),
\]
    we obtain the following four equations for the functions $A,\ C,\ T,\ U$:
\bearr
    \dot T^2 = m^2 \rho^2 b^2 + \kappa^2;                \label{5.2}
\inch
    \biggl(\frac{\dot A}{A} + 2\frac{\dot b}{b}\biggr)
                    \dot T + \ddot T = \xi;
\yyy
    h^{ij} \d_i U \d_j U = \kappa^2;                     \label{5.3}
\inch
    \frac{1}{C} h^{ij} \d_i C \d_j C + \Delta U = \xi,
\ear
    where $\kappa$ and $\xi$ are real separation constants.

    Let us first consider \eqs (5.2). Since for the function (5.1)
\[
    \dot T = -\frac{1}{|u_s|^2}, \cm   A = \frac{|u_s|^2}{b^2 \rho^2},
\]
    the second equation in (5.2) is satisfied if one puts $\xi=0$. Let us
    write down the first equation using the asymptotic expression (2.13):
\bear
     \frac{\left(1-|\lambda_s|^2\right) (s+1)
            \biggl[ 1 + \fracd{m^2\rho^2 b^2}{2(s+1)^2}
                   + o\biggl(\fracd{1}{s^2}\biggr)\biggr]}
     {\biggl\{1+|\lambda_s|^2
          + 2\biggl[ 1 + \fracd{m^2\rho^2 b^2}{2(s+1)^2}
            + o\biggl(\fracd{1}{s^2}\biggr)\biggr]
      \re \Bigl[ \lambda_s\,
                \e^{-2i\sO\eta}(1+o(1))\Bigr] \biggr\}^2}
        = \kappa^2 + (m\rho b)^2.                       \label{5.4}
\ear
    Neglecting all terms of orders smaller than $(s+1)^2$, we obtain
\[
     \kappa^2 = (s+1)^2 [1+ o(1)], \cm \lambda_s\to 0 \cm
            {\rm as} \quad s\to \infty.
\]
    Therefore the l.h.s. of (5.4) may be expanded in powers of $\lambda_s$:
\bearr
    (s+1)^2
    \biggl\{1+ \frac{m^2\rho^2 b^2}{(s+1)^2}
            + o\biggl(\frac{1}{s^2}\biggr)
    -4\biggl[1+ \frac{m^2\rho^2 b^2}{2(s+1)^2}
                    + o\biggl(\frac{1}{s^2}\biggr)
      \re \Bigl[ \lambda_s
             \e^{-2i\sO\eta}(1+o(1))\Bigr]+o(\lambda)\biggr\}
\nnnv  \inch \inch \inch \inch
    = (s+1)^2 [1+o(1)] + m^2\rho^2 b^2.
\earn
    This equation is satisfied only if \ $\kappa^2 = (s+1)^2 + o(1)$ \ and
\beq
    \lim\limits_{s\to\infty} (s+1)^2 \lambda_s =0.       \label{5.5}
\eeq
    One can also easily verify that there exist harmonic polynomials which
    exactly satisfy \eqs (5.3) with $\kappa^2 = (s+1)^2$ and $\xi=0$.

    Thus the correspondence principle formulated here selects among the Fock
    representations with $O(4)$-invariant quasi-vacua such representations
    that the sequences $\{\lambda_s\}$ characterizing them converge to zero
    faster than $\left\{(s+1)^{-2}\right\}$.

    Let us note that these selected representations are unitarily equivalent
    to each other. To prove that, it is sufficient to establish their
    unitary equivalence to the representation where all $\lambda_s = 0$. In
    quite a similar manner to the corresponding calculation in [5], it can
    be shown that this takes place if the product
\[
     \prod_{s=0}^{\infty} \left(1 - |\lambda_s|\right)^{-(s+1)^2/2}
\]
    converges. For its convergence, it is sufficient that
\[
    |\lambda_s| \leq \const\cdot s^{-(3/2 + \eps)}
\]
    where $\eps >0$ is an arbitrary number, and the set of sequences
    $\{\lambda_s\}$, selected by the correspondence principle, manifestly
    satisfies this requirement.

\twocol
\section{The limit of Minkowski space}

    Evidently, we should  also require that as $\rho\to\infty$, when the
    space-time curvature vanishes, the expression (4.3) for the field
    operator should turn into the usual decomposition into positive- and
    negative-frequency exponentials.

    To consider this limiting transition, we introduce the coordinates
\beq
     t=\rho\eta,\qquad  y^1 = \frac{r}{\rho}, \qquad y^2 =\theta,
                        \qquad  y^3 = \chi,
\eeq
    so that
\bear
    n^1 \eql \sin\frac{r}{\rho} \sin\theta \cos\chi,
                    \cm n^3 = \sin\frac{r}{\rho} \cos\theta,
\nn
    n^2 \eql \sin\frac{r}{\rho} \sin\theta \sin\chi,
                \cm n^4 = \cos \frac{r}{\rho}.
\ear
    Then
\[   \nq\,
    ds^2 = b^2\biggl(\frac{t}{\rho}\biggr)
    \biggl[ dt^2 - dr^2 -\rho^2 \sin^2\frac{r}{\rho}
                 (d\theta^2 + \sin^2\theta d\chi^2)\biggr],
\]
    which turns into the Minkowski metric in spherical coordinates in the
    limit $\rho\to\infty$:
\[
     ds^2 =  dt^2 - dr^2 - r^2 (d\theta^2 + \sin^2\theta d\chi^2).
\]

    The expansion (4.3) can be represented in these coordinates in the form
\bearr
     \phi(\eta,y) =
    		\frac{1}{\sqrt{2}b(\eta)\rho}
         		\sum_{s=0}^{\infty} \sum_{l=0}^{s} \sum_{n=-l}^{l}
\nnn \cm  \cm
      \left\{
        Y_{sln} (y) u_s(\eta) c^+_{sln} +
        Y^*_{sln} (y) u^*_s(\eta) c^-_{sln}  \right\},
\earn
    where
\bearr
     Y_{sln}(y) = \frac{\sqrt{\Gamma(s+l+3/2)\,(s+1)}}
                   {\sqrt{\Gamma(s-l+1/2) \sin(r/\rho)}}
\nnn \cm \cm \cm \times
           P^{-l+1/2}_{s+1/2} (\cos(r/\rho))\ Y_{ln}(\theta,\chi),
\earn
    $P\mN$ are Legendre functions and $Y_{ln}$ are normalized
    spherical functions.

    If we put
\[
    s/\rho = k_s, \qquad  1/\rho = \Delta k_s, \qquad
            \sqrt{\rho} c_{sln}^{\pm} = \phi_{ln}^{\pm}(k),
\]
    then
\bearr  \nq\,
      \phi = \frac{1}{\sqrt{2}} \sum_{l,n} \sum_{k_s}
                \Delta k_s
          \frac{\sqrt{\Gamma(s+l+3/2)}}{\sqrt{\Gamma(s-l+1/2}}
      \biggl(\sin\frac{r}{\rho}\biggr)^{-1/2}
\nnn \cm \cm \times
       P_{s+1/2}^{-(l+1/2)} \biggl(\cos \frac{k_s r}{s+1}\biggr)
            \sqrt{k_s}
\nnn \times
    \biggl\{ Y_{ln}(\theta,\chi) u_s\biggl(\frac{t}{\rho}\biggr)
            \phi^+_{ln} (k_s)
\nnn \inch \cm
           + Y^*_{ln}(\theta,\chi) u^*_s\biggl(\frac{t}{\rho}\biggr)
            \phi^-_{ln} (k_s)
            \biggr\}.
\earn
    Due to (2.10) and (2.12), the condition $b(0) =1$ and the well-known
    equality
\[
       \lim\limits_{\nu\to\infty} \nu^\mu
        P_{\nu}^{-\mu} \biggl(\cos \frac{x}{\nu}\biggr) =J_{\mu}(x),
\]
    where $J_\mu$ is a Bessel function, we obtain, as $\rho\to\infty$ and
    $k_s \to k > 0$,
\bearr \nq\,
      \phi(t,r,\theta,\chi) = \frac{1}{\sqrt{2}}
         \sum_{l=0}^{\infty} \sum_{n=-l}^{l} \intl_0^\infty dk\
\nnn \nhq \times
      \Biggl \{
        \frac{\e^{ik^0 t}}{\sqrt{k^0}}
                V_{ln} (k,y)\ \phi^+_{ln} (k) +
        \frac{\e^{-ik^0 t}}{\sqrt{k^0}}
                V^*_{ln} (k,y)\ \phi_{ln} (k) \Biggr\},
\nnn \cm
       k^0 = \sqrt{k^2 + m^2},
\nnn \cm
    V_{ln} (k,y) = \sqrt{\frac{k}{r}} J_{l+1/2}(kr)Y_{ln}(\theta,\chi),
\earn
    if $\lambda_s \to 0$ as $m\rho \to \infty$.

    A priori, we may not consider $\lambda_s$ to be independent of $m$ and
    $\rho$, but dimensional considerations imply that if such a dependence
    does exist, it should have the form $\lambda_s = \lambda_s(m\rho)$.

    The operators $\phi^{\pm}{}_{ln}(k)$ evidently obey the usual
    commutation rules for creation and annihilation operators
\bear
      \Bigl[\phi^+_{ln}(k),\ \phi^+_{l'n'}(k')\Bigr]  \eql 0,
\nn
      \Bigl[\phi^-_{ln}(k),\ \phi^+_{l'n'}(k')\Bigr]
                \eql \delta_{ll'} \delta_{nn'} \delta(k-k').
\earn

\section{Quasi-Euclidean isotropic model}

    Let us now briefly discuss a particular case of open models, the
    quasi-Euclidean isotropic Universe. Its metric can be represented
    in the form
\[
    ds^2 = a^2(\eta) [d\eta^2 -(d\xi^1)^2 -(d\xi^2)^2 -d\xi^3)^2]
\]
    and is invariant with respect to the group $O(3) \times T_3$. Solutions
    of the Klein-Gordon equation have the form
\bearr
    \phi(\eta,\ \vec \xi) = \frac{(2\pi)^{-3/2}}{a}
          \int u(\eta, \vec k) \e^{i\vec k \vec \xi} d\vec k,
\nnnv
    \vec k = (k_1,\ k_2,\ k_3); \quad\ \vec \xi = (\xi^1,\ \xi^2,\ \xi^3),
    \quad\   \vec k \vec \xi = k_i \xi^i.
\earn

    The quantum numbers $k_i$ are evidently eigenvalues of the generators
    $Z_i$ of the subgroup $T_3$. Changes in the formulae of \sect 2 are
    obvious, and we will not concentrate on them.

    Instead of (3.1), we now have
\bear
     z^-(\vec k) \eql \frac{1}{\sqrt{2}}
            \biggl\{p (\vec k) - \int d{\vec k{}'}
                T(\vec k,\vec k{}') q(\vec k{}')\biggr\},
\nn
     z^+(\vec k) \eql \left(z^-(\vec k)\right)^*, \cm
        T(\vec k, \vec k{}') = T(\vec k{}', \vec k),
\nnn
       \int \im T(\vec k, \vec k{}') f(\vec k) f(\vec k{}') d\vec k{}' > 0
\earn
     for any function $f(\vec k) \ne 0$.

     The condition of quasi-vacuum invariance with respect to $T_3$ gives
\[
       T(\vec k, \vec k{}') = T(\vec k) \delta(\vec k - \vec k{}'),
\]
     and with respect to $O(3)$:
\[
       T(\vec k) = T(k), \cm k = \sqrt{k_i k^i}.
\]
     Quasiclassical solutions at large $k$ exist if
\beq                                                           \label{7.1}
      k^2 \lambda(k) = k^2 \frac{i\sqrt{k^2+m^2} - T(k)}
                    {i\sqrt{k^2+m^2} + T(k)} \to 0
\eeq
     as $k \to \infty$.

     On the other hand, a representation of the commutation relations with a
     given function $T(k)$ is equivalent to the representation with
     $\lambda(k) = 0$ if $\int |\lambda(k)|^2\, k^2\, dk < \infty$,
     which is true if (7.1) is valid.

\section{Conclusion}

     We conclude that, unlike QFT in de Sitter space, in which the
     condition of quasi-vacuum invariance with respect to isometries and the
     correspondence principle single out the unique Fock representation
     $\{\lambda_s=0\}$, in our more general and consequently less symmetric
     case, it is apparently impossible to advance further than singling out
     a class of representations of the commutation relations which are
     equivalent to the one with all $\lambda_s =0$ or $\lambda(k) = 0$.
     However, the representations, satisfying all the requirements
     formulated in \sect 5 and 6, do not entirely fill this class. An
     interpretation of this fact probably requires an additional study.

     A few words on some special cases of the closed model. The case of
     de Sitter space ($b = 1/\cos \eta$) has been, as we already
     mentioned, considered in detail in Ref.\,[5]. The case of Einstein's
     static model ($b\equiv 1$) is the simplest. In this model, evidently,
\[
     u^{\pm}_s (\eta) = \frac{1}{\sqrt{\sO}}\e^{\pm i\sO\eta}.
\]
     The quantity $\MO$ is conserved for any rest mass $m$ and may be
     interpreted as the energy:
\[
     \MO = \Half \sum_{s,\sigma} \sqrt{m^2 + \frac{(s+1)^2}{\rho^2}}
             (c\s^+ c\s^- + c\s^- c\s^+).
\]
     The requirement (4.2) immediately leads to the condition $\lambda_s=0$.

     Furthermore, \eq (2.7) has solutions in terms of cylindrical functions
     for the metric obtained by Staniukovich [10] ($b = \e^{a\eta},\
     a=\const >0$) (in a theory of gravity with a time-variable ``constant''
     $\kappa$). The general solution (2.7) then has the form
\[
      u_s (\eta) = Z_{\pm i(s+1)/a}
                             \biggl(\frac{m\rho}{a} \e^{a\eta}\biggr),
\]
     where $Z_\nu (x)$ is any solution to the Bessel equation. The functions
     $u^{\pm}$ are then determined by the initial conditions (2.8).

     The authors are grateful to N.A. Chernikov for his attention and advice.

\noi
     The manuscript was received by JINR Publishing Division on 12 November
     1968.

\bigskip
\hrule
\bigskip

\section*{COMMENT}

The paper above is an English translation of an article written on the basis
of Dr. K. Bronnikov's graduation thesis at Moscow State University; the work
was carried out with my assistance. The paper was published in 1968 as
Joint Institute for Nuclear Research (Dubna) preprint P2--4151 in Russian.
For this reason it is not so internationally known as, e.g., Ref.\,[5], but
it had certain influence in the former USSR on studies of quantum fields and
particles interacting with gravitation, with further applications to
cosmology and astrophysics, which began in the second half of the 1960s.
Its results were essentially used in the first papers in this area by Ya.B.
Zel'dovich, A.A. Starobinsky, A.A. Grib and others. In particular, it is
known to me from a private communication by Prof. S.S. Gershtein that
Ya.B. Zel'dovich estimated it very highly. Probably, the paper can be not
only of historical interest, but also be useful for some present-day
readers.

\medskip

\rightline {\bf E.A. Tagirov}
\rightline {14 September 2004}

\end{document}